\documentclass[acmsmall,screen]{acmart} 
\renewcommand\footnotetextcopyrightpermission[1]{} 
\usepackage[english]{babel}
\usepackage[utf8x]{inputenc}
\usepackage[T1]{fontenc}
\usepackage{amsmath}
\usepackage{graphicx}
\usepackage{booktabs}
\usepackage{csquotes}
\usepackage{enumitem}
\usepackage{hyperref}
\usepackage{xcolor}
\usepackage{rotating}
\usepackage{microtype}        
\usepackage[all]{hypcap}      
\usepackage{ccicons}          
\usepackage[capitalize, noabbrev]{cleveref}
\usepackage[linecolor=red,backgroundcolor=yellow!40, textsize=scriptsize]{todonotes}
\usepackage{balance}
\makeatletter
\newcommand\footnoteref[1]{\protected@xdef\@thefnmark{\ref{#1}}\@footnotemark}
\makeatother

\newcommand{\x}[1]{#1}

\copyrightyear{2023} 
\acmYear{2023} 
\setcopyright{rightsretained} 
\acmConference[CHI '23]{Proceedings of the 2023 CHI Conference on Human Factors in Computing Systems}{April 23--28, 2023}{Hamburg, Germany}
\acmBooktitle{Proceedings of the 2023 CHI Conference on Human Factors in Computing Systems (CHI '23), April 23--28, 2023, Hamburg, Germany}
\acmDOI{10.1145/3544548.3581196}
\acmISBN{978-1-4503-9421-5/23/04}

\begin{document}
\pagestyle{plain}

\title{\x{Co-Writing} with Opinionated Language Models \x{Affects} Users' Views}
\renewcommand{\shorttitle}{Writing with Opinionated Language Models \x{Affects} Users' Views}

\begin{abstract}
If large language models like GPT-3 preferably produce a particular point of view, they may influence people's opinions on an unknown scale. This study investigates whether a \x{language-model-powered writing assistant that generates some opinions more often than others impacts what users write -- and what they think. In an online experiment, we asked participants (N=1,506) to write a post discussing whether social media is good for society. Treatment group participants used a language-model-powered writing assistant configured to argue that social media is good or bad for society. Participants then completed a social media attitude survey, and independent judges (N=500) evaluated the opinions expressed in their writing. Using the opinionated language model affected the opinions expressed in participants' writing and shifted their opinions in the subsequent attitude survey. We discuss the wider implications of our results} and argue that the opinions built into AI language technologies need to be monitored and engineered more carefully.
\end{abstract}

\author{Maurice Jakesch}
\affiliation{%
  \institution{Cornell University}
  \city{Ithaca}
  \state{New York}
  \country{USA}
}
\email{mpj32@cornell.edu}

\author{Advait Bhat}
\affiliation{
  \institution{Microsoft Research}
  \city{Bengaluru}
  \country{India}
}

\author{Daniel Buschek}
\affiliation{
  \institution{University of Bayreuth}
  \city{Bayreuth}
  \country{Germany}
}

\author{Lior Zalmanson}
\affiliation{
  \institution{Tel Aviv University}
  \city{Tel Aviv}
  \country{Israel}
}

\author{Mor Naaman}
\affiliation{%
  \institution{Cornell Tech}
  \city{New York}
  \state{New York}
  \country{USA}
}

\renewcommand{\shortauthors}{Jakesch et al.}

\maketitle

\section{Introduction}

\begin{figure*}
  \begin{center}
    \includegraphics[width=0.75\textwidth, trim=3.5cm 3.5cm 6cm 3.2cm, ,clip]{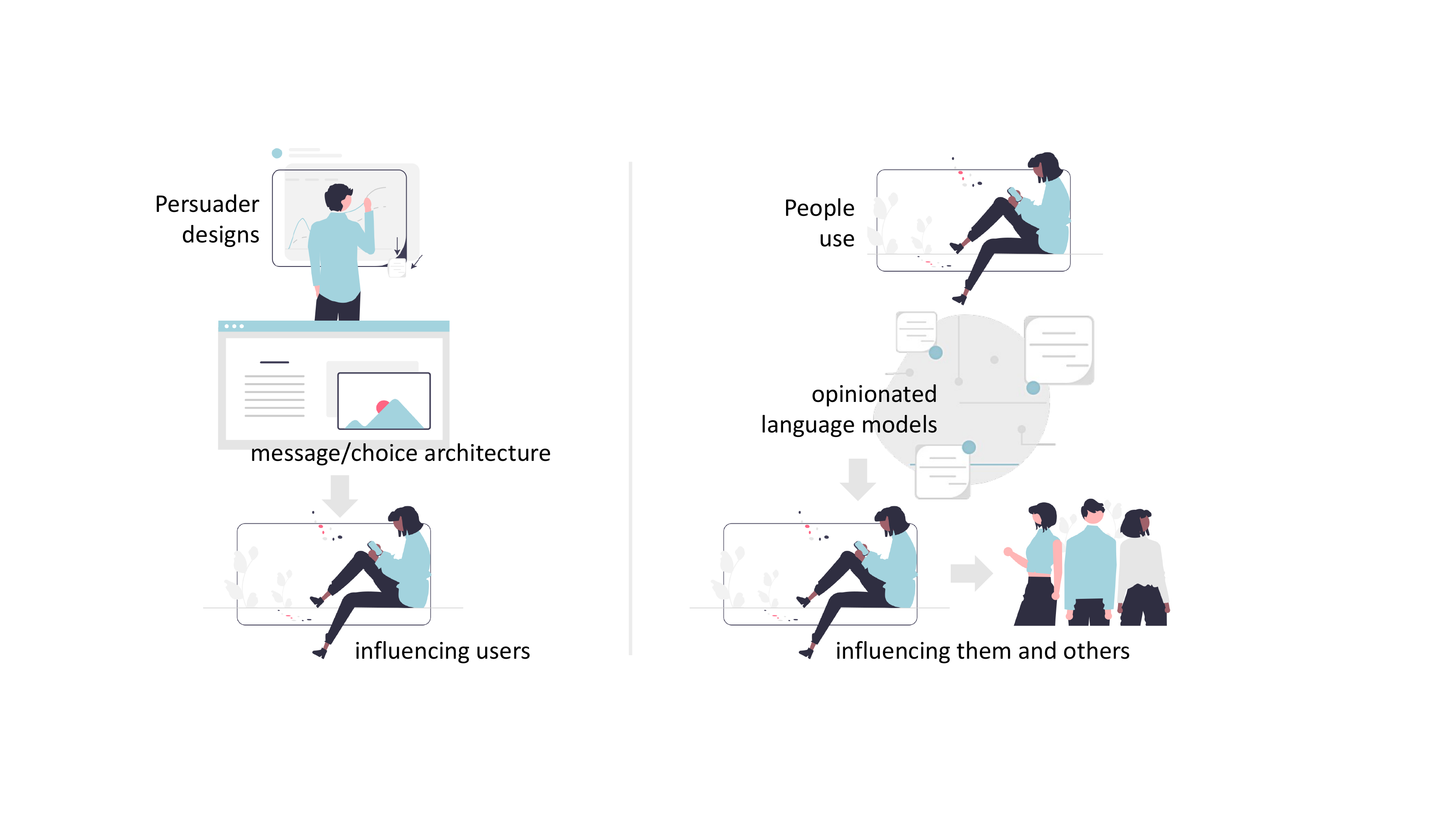}
  \end{center}
\caption{\textbf{Conventional technology-mediated persuasion (left) compared to \emph{latent persuasion} by language models (right).} In conventional influence campaigns, a central persuader designs an influential message or choice architecture distributed to recipients. In \emph{latent persuasion}, language models produce some opinions more often than others, influencing what their users write, which is, in turn, read by others.}
\label{fig:schema}
\end{figure*}

\x{Large language models like GPT-3~\cite{winata2021language, bommasani2021opportunities, vaswani2017attention} are increasingly becoming part of human communication. Enabled by developments in computer hardware and software architecture~\cite{vaswani2017attention}, large language models produce human-like language \cite{jakesch2022human} by iteratively predicting likely next words based on the sequence of preceding words. Applications like writing assistants \cite{dang2022beyond}, grammar support \cite{koltovskaia2020student}, and machine translation \cite{gaspari2014perception} inject the models' output into what people write and read \cite{Hancock2020}.}

Using large language models \x{in our daily communication may} change how we form opinions and influence each other. 
In conventional forms of persuasion, a persuader crafts a compelling message and delivers it to recipients -- either face-to-face or mediated through contemporary technology~\cite{simons2011persuasion}. More recently, user researchers and behavioral economists have shown that technical choice architectures, \x{such as the order of options presented} affect people's behavior as well~\cite{leonard2008richard, fogg2002persuasive}.
With the emergence of large language models that produce human-like language~\cite{jakesch2022human, buchanan2021truth}, interactions with technology may \x{influence} not only behavior but also opinions: when language models produce some views more often than others, they may persuade their users. We call this new paradigm of influence \emph{latent persuasion} by language models, illustrated in Figure \ref{fig:schema}.

\emph{Latent persuasion} by language models extends the insight that choice defaults \x{affect} people's behavior  \cite{leonard2008richard, fogg2002persuasive} to the field of language and persuasion. Where \emph{nudges} change behavior by making some \x{choices more convenient} than others, AI language technologies may shift opinions by making it easy to express certain views but not others. \x{Such influence could be \emph{latent} and hard to pinpoint:} choice architectures are visible, but opinion preferences built into language models may be opaque to users, policymakers, and even system developers. While in traditional persuasion, a central designer \x{intentionally creates a message to convince a specific audience, a language model may be opinionated by accident and its opinions may vary by user, product and context}.

Prior research on the risks of generative language models has focused on conventional persuasion scenarios, where \x{a human persuader uses} language models to automate and optimize the production of content for advertising ~\cite{sunny, duerr2021persuasive} or misinformation~\cite{kreps2022all, buchanan2021truth, zellers2019defending}.
\x{Initial audits also highlight} that language models reproduce stereotypes and biases~\cite{huang2019reducing, brown2020language, nozza2021honest} and support certain cultural values more than others~\cite{johnson2022ghost}. While emerging research on co-writing with large language models suggests that models become increasingly active partners in people's writing~\cite{Lee2022coauthor, Yang2022AIAA, Yuan2022wordcraft}, little is known about how the opinions produced by language models affect users' views.
Work by~\citet{Arnold2018review_sentiment} and \citet{Bhat2021, Bhat2022} shows that a biased writing assistant may affect movie or restaurant reviews, but whether \x{co-writing with large language models} affect users' opinions on public issues remains an open and urgent question.

This study investigates whether large language models that generate certain opinions more often than others \x{affect} what their users write and think.
In an online experiment \x{(N=1,506)}, participants wrote a short statement discussing whether social media is good or bad for society. Treatment group participants were shown suggested text generated by a large language model. The model, GPT-3~\cite{winata2021language}, was configured to either generate text that argued that social media is good for society or text that argued the opposite. Following the writing task, we asked participants to assess social media's societal impact in a survey.
A separate sample of human judges (N=500) evaluated the opinions expressed in participants' writing.

Our quantitative analysis tests whether the interactions with the opinionated language model \x{shifted} participants' writing and survey opinions. We explore how this opinion \x{shift} may have occurred \x{in secondary analyses}. We find that both \x{participants' writing} \textit{and} their attitude towards social media  in the survey were considerably \x{affected} by the model's preferred opinion. We conclude by discussing how \x{researchers, AI practitioners, and policymakers can respond to the possibility of latent persuasion by AI language technologies.}

\section{Related work}
\x{Our study is informed by} prior research on social influence and persuasion, interactions with writing assistants, and the societal risks of large language models.

\subsection{Social influence and persuasion}
Social influence is defined as a shift in an individual's thoughts, feelings, attitudes, or behaviors as a result of interaction with others~\cite{rashotte2007social}. While social influence is integral to human collaboration, landmark studies have shown that it can also lead to unreasonable or unethical behavior. On a personal level, people may conform to majority views against their better judgement~\cite{asch1951effects} and obey an authority figure even if it means harming others~\cite{milgram1963behavioral}. On a societal level, researchers have shown that social influence drives speculative markets~\cite{shiller2015irrational}, affects voting patterns~\cite{lazarsfeld1968people}, and contributes to the spread of unhealthy behaviors such as smoking and obesity~\cite{christakis2007spread, christakis2008collective}.

Following the rise of social media, how online interactions affect people's opinions and decisions has been studied extensively. 
Research has shown that a variety of sources influences users' attitudes and behaviors, including friends, family, experts, and internet celebrities~\cite{goel2012structure,marwick2011tweet}; the latter group was labeled \emph{influencers} due to their influence on a large group of 'followers'~\cite{bakshy2011everyone}. Research has  found that in online settings, users can be influenced by non-human entities such as brand pages, bots, and algorithms~\cite{ferrara2016rise}. Studies have evaluated the influence that technical artifacts such as personalized recommendations, chatbots, and choice architectures have on users' decision-making~\cite{berkovsky2012influencing, leonard2008richard, cosley2003seeing, gunaratne2018persuasive}. 

The influence that algorithmic entities have on people depends on how people perceive the algorithm, for example, whether they attribute trustworthiness to its recommendations~\cite{logg2019algorithm, gunaratne2018persuasive}. The influence of algorithms on individuals tends to increase as the environment becomes more uncertain and decisions become more difficult~\cite{bogert2021humans}. 
With the public's growing awareness of developments in artificial intelligence, people may regard \emph{smart} algorithms as a source of authority ~\cite{kapania2022because, logg2019algorithm, araujo2020ai}. There is recent evidence that people may accept algorithmic advice even in simple cases when it is clearly wrong~\cite{liel2020if}. In the related field of automation, such over-reliance on machine output has been referred to as \emph{automation bias}~\cite{parasuraman1997humans,parasuraman2010complacency, wickens2015complacency}. 

\subsection{Interaction with writing assistants}

Historically, HCI research for text entry has predominantly focused on efficiency~\cite{Kristensson2014}. Typical text entry systems attend to language context at the word~\cite{Vertanen2015velocitap, Bi2014} or sentence level~\cite{Arnold2016phrases_vs_words, Buschek2021emails}. They suggest one to three subsequent words based on underlying likelihood distributions~\cite{Dunlop2012, Fowler2015, Gordon2016, Quinn2016chi}. More recent systems also provide multiple short reply suggestions~\cite{Kannan2016smartreply} or a single long phrase suggestion~\cite{Chen2019smartcompose}.
More extensive suggestions have traditionally been avoided because the time taken to read and select them might exceed the time required to enter that text manually. Several studies indicate that features such as auto-correction and word suggestions can negatively impact typing performance and user experience~\cite{Banovic2019mobilehci, Dalvi2016,  Buschek2018researchime, Palin2019}.

However, with the emergence of larger and more powerful language models \cite{winata2021language, bommasani2021opportunities, vaswani2017attention},  there has been a growing interest in design goals beyond efficiency. Studies have investigated interface design factors and interactions with writing assistants that directly or indirectly support inspiration~\cite{Lee2022coauthor, Singh2022elephant, Yuan2022wordcraft, Bhat2022}, language proficiency~\cite{Buschek2021emails}, story writing~\cite{Singh2022elephant, Yuan2022wordcraft}, text revision~\cite{Cui2020, Zhang2019} or creative writing~\cite{Clark2018, Gero2019MetaphoriaAA}. Here, language models are framed as \textit{active writing partners} or \textit{co-authors}~\cite{Lee2022coauthor, Yang2022AIAA, Yuan2022wordcraft}, rather than tools for prediction or correction. There is  evidence that a system that suggests phrases rather than words~\cite{Arnold2016phrases_vs_words} is more likely to be perceived as a collaborator and content contributor by users.

The more writing assistants become \textit{active writing partners} rather than mere tools for text entry, the more the writing process and output may be affected by their ``co-authorship''.
\citet{Bhat2022} discuss how writers evaluate the suggestions provided and integrate them into different cognitive writing processes. Work by \citet{Singh2022elephant} suggests that writers make active efforts or 'leaps' to integrate generated language into their writing.
\citet{Buschek2021emails} conceptualized nine behavior patterns that indicate varying degrees of engagement with suggestions, from ignoring them to chaining multiple ones in a row.
Writing with suggestions correlates with shorter and more predictable texts being written~\cite{Arnold2020image_captions}, along with increased use of standard phrases when writing with a language model~\cite{Buschek2021emails, Bhat2022}. Furthermore, the sentiment of the suggested text may \x{affect} the sentiment of the written text~\cite{Arnold2018review_sentiment, Hohenstein2020}. 

\subsection{Societal risks of large language models}
Technical advances have given rise to a generation of language models~\cite{bommasani2021opportunities} that produces language so natural that humans can barely distinguish it from real human language~\cite{jakesch2022human}. Enabled by improvements in computer hardware and the transformer architecture~\cite{vaswani2017attention}, models like GPT-3~\cite{brown2020language, radford2019language} have attracted attention for their potential to impact a range of beneficial real-world applications~\cite{bommasani2021opportunities}. However, more cautious voices have also warned about the ethical and social risks of harm from large language models~\cite{weidinger2021ethical, weidinger2022taxonomy}, ranging from discrimination and exclusion~\cite{huang2019reducing, brown2020language, nozza2021honest} to misinformation~\cite{kreps2022all, lin2021truthfulqa, rae2021scaling, zellers2019defending} and environmental~\cite{strubell2019energy} and socioeconomic harms~\cite{bender2021dangers}.

Comparatively little research has considered widespread shifts in opinion, attitude, and culture that may result from a comprehensive deployment of generative language models. It is known that language models work and perform better for the languages and contexts they are trained in (primarily English or Mandarin Chinese)~\cite{brown2020language, rae2021scaling, winata2021language}. Small-n audits have also suggested that the values embedded in models like GPT-3 were more aligned with reported dominant US values than those upheld in other cultures~\cite{johnson2022ghost}. Work by \citet{jakesch2022different} has highlighted that the values held by those developing AI systems differ from those of the broader populations interacting with the systems.
The adjacent question of AI alignment -- how to build AI systems that act in line with their operators' goals and values -- has received comparatively more attention~\cite{askell2021general}, but primarily from a control and safety angle. 

A related topic, the political repercussions of social media and recommender systems~\cite{zhuravskaya2020political}, has received extensive research attention, however. After initial excitement about social media's democratic potential~\cite{khondker2011role}, it became evident that technologies that affect public opinion will be subject to powerful political and commercial interests~\cite{bradshaw2017troops}. Rather than mere technical platforms, algorithms become constitutive features of public life~\cite{gillespie2014relevance} that may change the political landscape~\cite{aral2019protecting}. Even without being designed to \x{shift} opinions, it has been found that algorithms may contribute to political polarization by reinforcing divisive opinions~\cite{bruns2019filter, cinelli2021echo, bail2018exposure}. 
\section{Methods}
To investigate whether interacting with opinionated language models shifts people's writing and affects people's views, we conducted an online experiment asking participants \x{(N=1,506)} to respond to a social media post in a simulated online discussion using a writing assistant. The language model powering this writing assistant was configured to generate text supporting one or the other side of the argument. We compared the essays and opinions of participants to a control group that wrote their social media posts without writing assistance.

\begin{figure*}
  \begin{center}
    \includegraphics[width=0.87\textwidth, , trim=1cm 2.2cm 1cm 1cm]{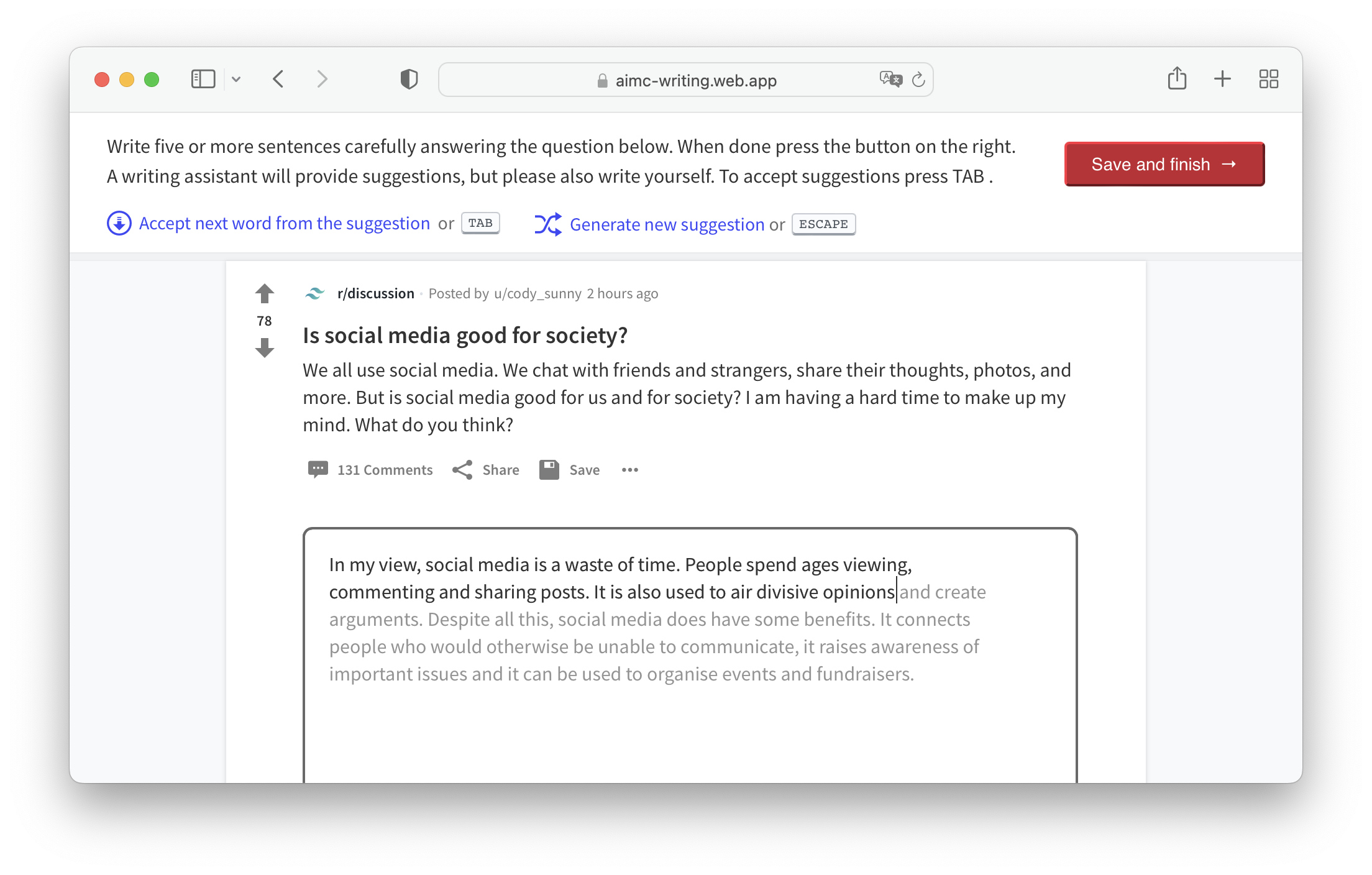}
  \end{center}
\caption{\textbf{Screenshot of the writing task.} The task is shown on the top of the page, followed by usage instructions for the writing assistant. Below, participants read a Reddit-style discussion post to which they were asked to reply. The writing assistant displayed writing suggestions (shown in grey) extending participants' text. The participant in the screenshot wrote an argument critical of social media, but the model is configured to argue that social media is \textit{good} for society.}
\label{fig:screenshot}
\end{figure*}

\subsection{Experiment design}
To study interactions between model opinion and participant opinion in a possibly realistic and relevant setting, we created the scenario of an opinionated discussion on social media platforms like Reddit. Such discussions have a large readership~\cite{medvedev2017anatomy}, pertain to political controversies, and are plausible application settings for writing assistants and language models. We searched ProCon.org\footnote{\url{https://www.procon.org/}}, an online resource for research on controversial issues, to identify a discussion topic. We selected ``Is Social Media Good for Society?'' as a discussion topic. We chose this topic because it is an easily accessible discussion topic that is politically relevant but not considered so controversial that entrenched views may limit constructive debate. 

To run the experiment, we created a custom experimental platform combining a mock-up of a social media discussion page, a rich-text editor, and a writing assistant. The assistant was powered by a language generation server and included comprehensive logging tools. 
To provide a realistic-looking social media mock-up, we copied the design of a Reddit discussion page and drafted a question based on the ProCon.org discussion topic. Figure~\ref{fig:screenshot} shows a screenshot of the experiment. We asked participants to write at least five sentences expressing their take on social media's societal impact.
We randomly assigned participants to three different treatment groups:
\begin{enumerate}
\item \emph{Control group:} participants wrote their answers without a writing assistant.
\item \emph{Techno-optimist language model treatment:} participants were shown suggestions from a language model configured to argue that social media is good for society.
\item \emph{Techno-pessimist language model treatment:} participants received suggestions from a language model configured to argue that social media is bad for society.
\end{enumerate}

\subsection{Building the writing assistant}
Similar to Google's \textit{Smart Compose}~\cite{Chen2019smartcompose} and Microsoft's predictive text in Outlook, the writing assistant in the treatment groups suggested possible continuations (sometimes called ``completions'') to text that participants had entered. 
We integrated the suggestions into a customized version of the rich-text editor Quill.js\footnote{\url{https://quilljs.com/}}. The client sent a generation request to the server whenever a participant paused their writing for a certain amount of time (750ms). Including round-trip and generation time, a suggestion appeared on participants' screens about 1.5 seconds after they paused their writing.

When the editor client received a text suggestion from the server, it revealed the suggestion letter by letter with random delays calibrated to resemble a co-writing process (cf.~\cite{Lehmann2022muc}). Once the end of a suggested sentence was reached, the editor would pause and request from the server an extended generation until at least two sentences had been suggested. Participants could accept each suggested word by pressing the tab key or clicking an accept button on the interface. In addition, they could reset the generation, requesting a new suggestion by pressing a button or key. 

We hosted the required cloud functions, files, and interaction logs on Google's Firebase platform. 

\subsection{Configuring an opinionated language model}
In this study, we experimented with language models that \textit{strongly} favored one view over another. We chose a strong manipulation as we wanted to explore the \textit{potential} of language models to affect users' opinions and understand whether they could be used or abused to shift people's views~\cite{bagdasaryan2021spinning}.

We used GPT-3~\cite{brown2020language} with manually designed prompts to generate text suggestions for the experiment in real-time. Specifically, we accessed OpenAI's most potent 175B parameter model (``text-davinci-002''). \x{We used temperature sampling, a method for choosing a specific next token from the set of likely next tokens inspired by statistical thermodynamics. We set the sampling temperature (randomness parameter) to 0.85 to generate suggestions that are  varied and creative. We set the frequency and presence penalty parameters to 1 to reduce the chance that the model suggestions would become repetitive. We also prevented the model from producing new lines, placeholders, and list by setting logit bias parameters that reduced the likelihood of the respective tokens being selected. }

\x{We evaluated different techniques to create an opinionated model, i.e., a model that \emph{likely supports a certain side of the debate} when generating a suggestion. We used prompt design \cite{lester_constant_2022}, a technique for guiding frozen language models to perform a specific task. Rather than updating the weights of the underlying model, we concatenated an engineered prompt to the input text to increase the chance that the model generates a certain opinion. Specificially we inserted the prefix} \emph{"Is social media good for society? {Explain why social media is good/bad for society:}"} before participants' written texts when generating continuation suggestions. \x{The engineered prompt was not visible to participants in their editor UI; it was inserted in the backend before generation and removed from the generated text before showing it to participants. }

\x{Initial experimentation and validation suggested that the  prompt  produced the desired opinion in the generated text, but when participants  strongly argued for another opinion in their writing, the model's continuations would follow their opinion. In addition to the prefix prompt, we thus developed an infix prompt that would be inserted throughout participants' writing to reinforce the desired opinion.} We inserted the snippet   (\emph{"{One sentence continuing the essay explaining why social media is good/bad:}"}) \x{right before the last sentence that participants had written. This additional prompt guided the model's continuation towards the target opinion even if participants had articulated a different opinion earlier in their writing. }
Validation of the model opinion configuration is provided in section~\ref{sec:validation}. We also experimented with fine-tuning~\cite{howard2018universal} \x{to guide the models' opinion, but the fine-tuned models} did not consistently produce the intended opinion.

\subsection{Outcome measures and covariates}
We collected different types of outcome measures to investigate interactions between participants' opinions and the model opinion:

\emph{Opinion expressed in the post:} To evaluate expressed opinion, we split participants' written texts into sentences and asked crowd workers to evaluate the opinion expressed in each sentence. 
Each crowd worker assessed 25~sentences, indicating whether each argued that social media is good for society, bad, or both good and bad. A fourth label was offered for sentences that argued neither or were unrelated. For example, \emph{"Social media also promotes cyber bullying which has led to an increase in suicides" (P\#421)} was labeled as arguing that social media is bad for society, while \emph{"Social media also helps to create a sense of community" (P\#1169)} was labeled as \emph{social media is good for society}. We collected one to two labels for each sentence participants wrote and collected labels for a sample of the writing assistant's suggestions. In sentences where we collected multiple labels, the labels provided by different raters agreed 84.1\% of the time (Cohen's $\kappa=0.76$).

\emph{Real-time writing interaction data:} We gathered comprehensive interaction logs at the key-stroke level of how participants interacted with the model's suggestions. We recorded which text the participant had written, what text the model had suggested, and what suggestions participants had accepted from the writing assistant. We measured how long they paused to consider suggestions and how many suggestions they accepted. 

\emph{Opinion survey (post-task):} After finishing the writing task, participants completed an opinion survey. The central question, ``Overall, would you say social media is good for society?'' was designed to assess \x{shifts} in participants' attitude. This question was not shown immediately after the writing task to reduce demand effects. 
Secondary questions were asked to understand participants' opinions in more detail: ``How does social media affect your relationships with friends and family?'', ``Does social media usage lead to mental health problems or addiction?'', ``Does social media contribute to the spread of false information and hate?'', ``Do you support or oppose government regulation of social media companies?'' 
The questions were partially adapted from Morning Consults' National Tracking Poll~\cite{consultnational}; answers were given on typical 3- and 5-point Likert scales.

\emph{User experience survey (post-task):} Participants in the treatment groups completed a survey about their experience with the writing assistant following the opinion survey. They were asked, ``How useful was the writing assistant to you?'', whether ``The writing assistant understood what you wanted to say'' and whether ``The writing assistant was knowledgeable and had expertise.'' To explore participants' awareness of the writing assistant's opinion and \x{its effect on their own views}, we asked them whether ``The writing assistant's suggestions were reasonable and balanced'' and whether ``The writing assistant inspired or changed my thinking and argument.'' Answers were given on a 5-point Likert scale from ``strongly agree'' to ``strongly disagree.'' An open-ended question asked participants what they found most useful or frustrating about the writing assistant.

\emph{Covariates:} We asked participants to self-report their age, gender, political leaning, and their highest level of education at the end of the study. We also constructed a ``model alignment'' covariate estimating whether the opinion the model supported was aligned with the participant's opinion. We did not ask participants to report their overall judgment before the writing task to avoid commitment effects. Instead, we asked them at the end of the study whether they believed social media was good for society before participating in the discussion. While imperfect, this provides a proxy for participants' pre-task opinions. It is biased by the treatment effect observed on this covariate, which amounts to 14\% of its standard deviation.

\subsection{Participant recruitment}
We recruited 1,506 participants (post-exclusion) for the writing task, corresponding to 507, 508, and 491 individuals in the control, techno-optimist, and techno-pessimist treatment groups, respectively. The sample size was calculated based on effect sizes observed in the pilot studies' post-task question, "Overall, would you say social media is good for society?" at a power of 80\%. The sample was recruited through Prolific~\cite{palan2018prolific}. The sample included US-based participants at least 18 years old (M= 37.7, SD= 14.2); 48.5\% self-identified as female, and 48.6\% identified as male. \x{38 participants identified as non-binary and eight preferred to self-describe or not disclose their gender identity. }
Six out of ten indicated liberal leanings; 57.1\% had received at least a Bachelor's degree. Participants who failed the pre-task attention check (8\%) were excluded. Six percent of participants admitted to the task did not finish it. We paid participants \$1.50 for an average task time of 5.9 minutes based on an hourly compensation rate of \$15. For the labeling task, we recruited a similar sample of 500 participants through Prolific. The experimental protocols were approved by the Cornell University Institutional Review Board.

\subsection{Data sharing}
\x{The experiment materials, analysis code and data collected are publicly available through an Open Science repository (\url{https://osf.io/upgqw/}}). A research assistant screened the data, and records with potentially privacy-sensitive information were removed before publication.

\section{Results}
We first analyze the opinions participants expressed in their social media posts. We then examine whether participants may have accepted the models' suggestions out of mere convenience and whether the model influenced participants' opinions in a later survey. Finally, we present data on participants' perceptions of the model's opinion and influence. The reported statistics are based on a logistic regression model.

\subsection{Did the interactions with the language model affect participants' writing?}

\begin{figure*}
  \begin{center}
    \includegraphics[width=0.82\textwidth, trim=0cm 0cm 0cm 0cm]{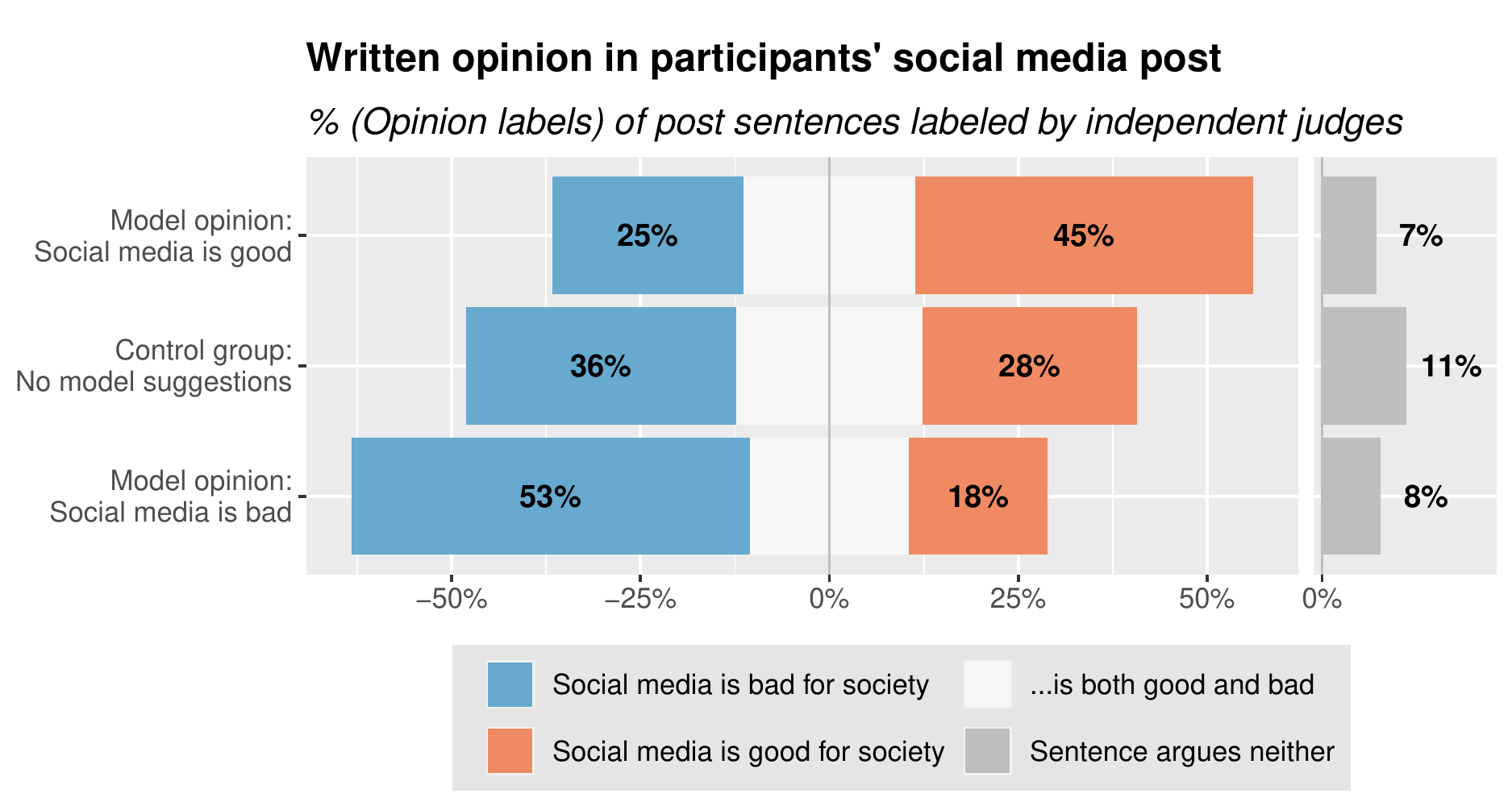}
  \end{center}
\caption{\textbf{Participants assisted by a model supportive of social media were more likely to argue that social media is good for society in their posts (and vice versa).} N\textsubscript{s}=9,223 sentences written by N\textsubscript{p}=\x{1,506} participants evaluated by N\textsubscript{j}=500 judges. The y-axis indicates whether participants wrote their social media posts with assistance from an opinionated language model that was supportive (top) or critical of social media (bottom). The x-axis shows how often participants argued that social media is bad for society (blue), good for society (orange), or both good and bad (white) in their writing.}
\label{fig:written-opinion}
\end{figure*}

Figure~\ref{fig:written-opinion} shows how often participants in each of the treatment conditions (y-axis) argued that social media is good or bad for society (x-axis) in their writing. The social media posts written by participants in the control group (middle row) were slightly critical of social media: They argued that social media is bad for society in 38\% and that social media is good in 28\% of their sentences. In about 28\% of their sentences, control group participants argued that social media is both good and bad, and 11\% of their sentences argued neither or were unrelated. 

Participants who received suggestions from a language model supportive of social media (top row of Figure~\ref{fig:written-opinion}) were 2.04 times more likely than control group participants (p<0.0001, 95\% CI [1.83, 2.30]) to argue that social media is good. In contrast, participants who received suggestions from a language model that criticized social media (bottom row) were 2.0 times more likely (p<0.0001, 95\% CI [1.79, 2.24] to argue that social media is bad than control group participants. 
We conclude that using an opinionated language model \x{affected} participants' writing such that the text they wrote was more likely to support the model's preferred view.

\subsection{Did participants accept the model's suggestions out of mere convenience?}
Participants may have accepted the models' suggestions out of convenience, even though the suggestions did not match what they would have wanted to say. Paid participants in online studies, in particular, may be motivated to accept suggestions to swiftly complete the task. 

\begin{figure*}
  \begin{center}
    \includegraphics[width=0.82\textwidth, trim=0cm 0cm 0cm 0cm]{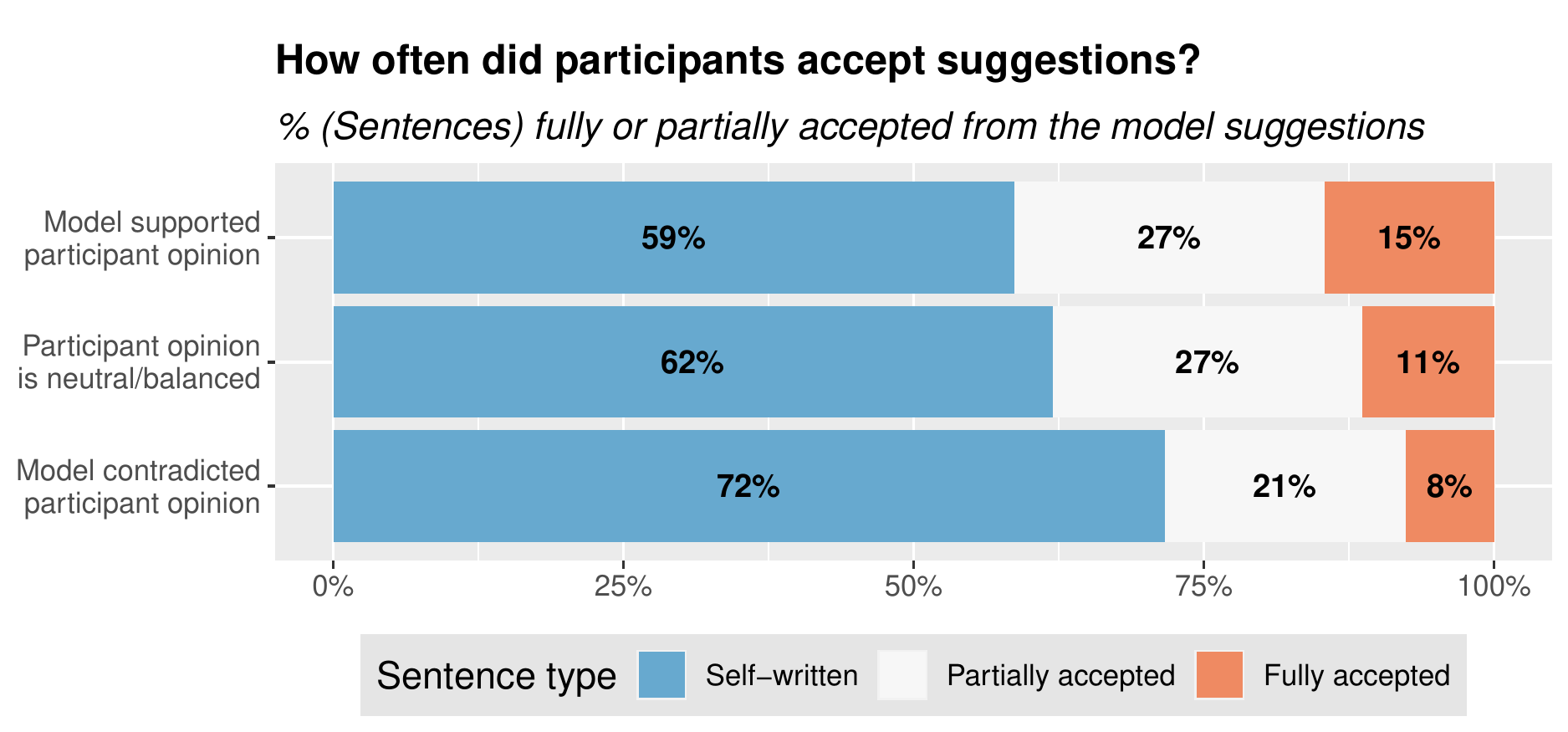}
  \end{center}
\caption{\textbf{Participants were more likely to accept suggestions if the model's opinion aligned with their own views} N\textsubscript{s}=6,142 sentences by N\textsubscript{p}=1,000 participants. The x-axis shows how many of the sentences participants had written themselves (blue), together with the model (white), or fully accepted from the model's suggestions (orange). The y-axis disaggregates the data based on whether the model suggestions were in line with participants' likely pre-task opinion.}
\label{fig:acceptance-ratio}
\end{figure*}

Our data shows that, across conditions and treatments, most participants did not blindly accept the model's suggestions but interacted with the model to co-write their social media posts. 
On average, participants wrote 63\% of their sentences themselves without accepting suggestions from the model (compare Figure~\ref{fig:acceptance-ratio}). About 25\% of participants' sentences were written by both the participant and the model, which typically meant that the participant wrote some words and accepted the model's remaining sentence suggestion. Only 11.5\% of sentences were fully accepted from the model. Participants whose personal views were likely aligned with the model were more likely to accept suggestions, while participants with opposing views accepted fewer suggestions. About one in four participants did not accept any model suggestion, and one in ten participants had more than 75\% of their post written by the model.

\begin{figure*}
  \begin{center}
    \includegraphics[width=0.82\textwidth, trim=0cm 0cm 0cm 0cm]{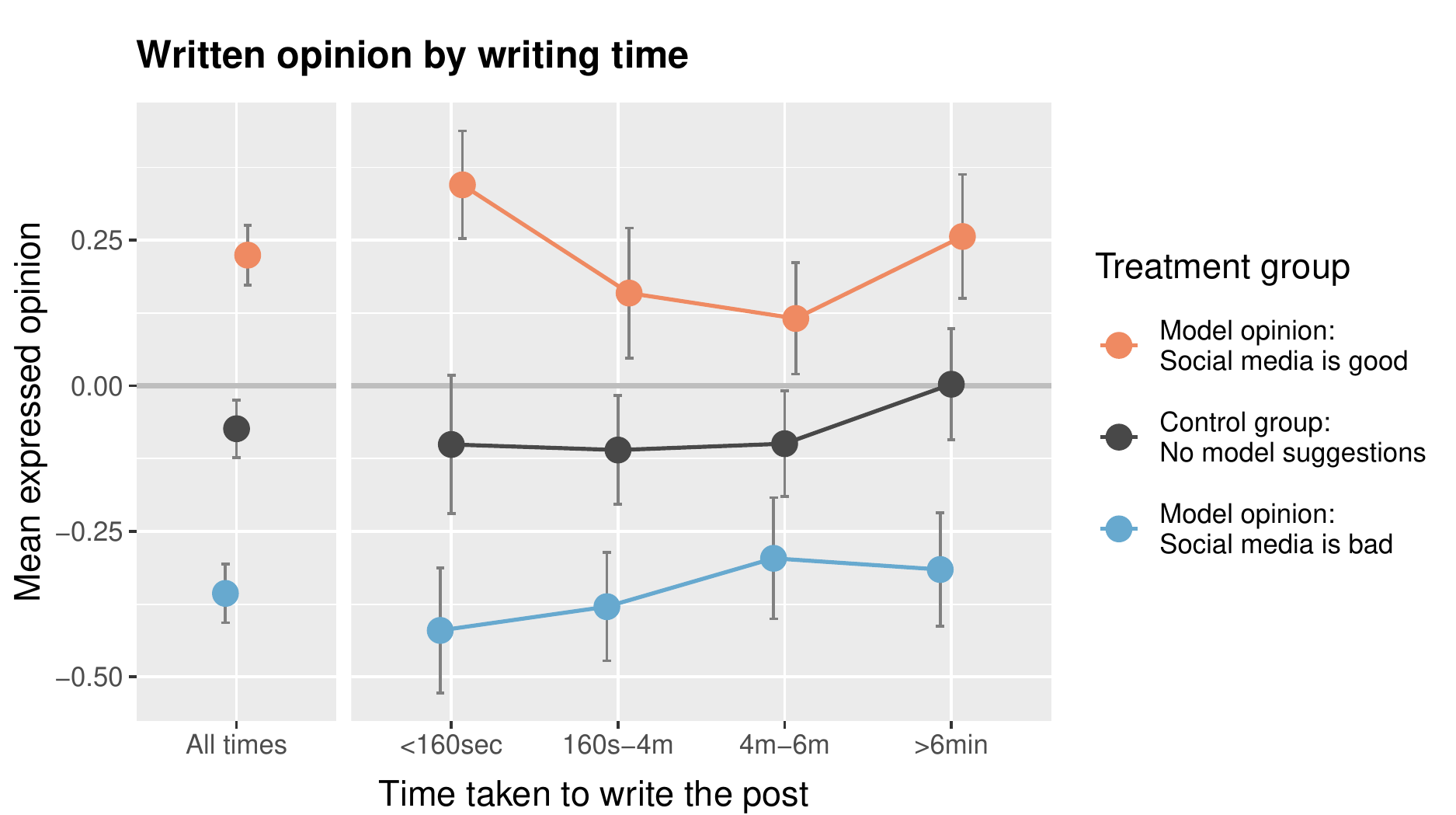}
  \end{center}
\caption{\textbf{The opinion \x{differences} in participants' writing were larger when they finished the task quickly.} \x{N=1,506}. The y-axis shows the mean opinion expressed in participants' social media posts based on aggregated sentence labels ranging from -1 for ``social media is bad for society'' to 1 for ``social media is good for society''. The x-axis indicates how much time participants took to write their posts. For reference, the left panel shows expressed opinions aggregated across writing times.}
\label{fig:opinion-time}
\end{figure*}

\subsubsection{Did conveniently accepted suggestions increase the observed differences in written opinion?}
The writing of participants who spent little time on the task was more \x{affected} by the model's opinion. We use the time participants took to write their posts to estimate to what extent they may have accepted suggestions without due consideration. For a concise statistical analysis, we treat the ordinal opinion scale as an interval scale. Since the opinion scale has comparable-size intervals and a zero point, continuous analysis is meaningful and justifiable~\cite{knapp1990treating}. We treat ``social media is bad for society'' as -1 and ``social media is good for society'' as 1. Sentences that argue both or neither are treated as zeros. 

Figure~\ref{fig:opinion-time} shows the mean opinion expressed in participants' social media posts depending on treatment group and writing time. The left panel shows participants' expressed opinions across times for reference, with a mean opinion difference of about 0.29 (p<0.001, 95\% CI [0.25, 0.33], SD=0.58) between each treatment group and the control group (corresponding to a large effect size of d=0.5). Participants who took little time to write them (less than 160 seconds, left-most data in right panel) were more affected by the opinion of the language model (0.38, p<0.001, 95\% CI [0.31, 0.45]). Our analysis shows that accepting suggestions out of convenience has contributed to the differences in the written opinion. However, even for participants who took four to six minutes to write their posts, we observed significant differences in opinions across treatment groups (0.20, p<0.001, 95\% CI [0.13, 0.27], corresponding to a treatment effect of d=0.34). 

\subsection{Did the language model affect participants' opinions in the attitude survey?}

\begin{figure*}
  \begin{center}
    \includegraphics[width=0.82\textwidth, trim=0cm 0cm 0cm 0cm]{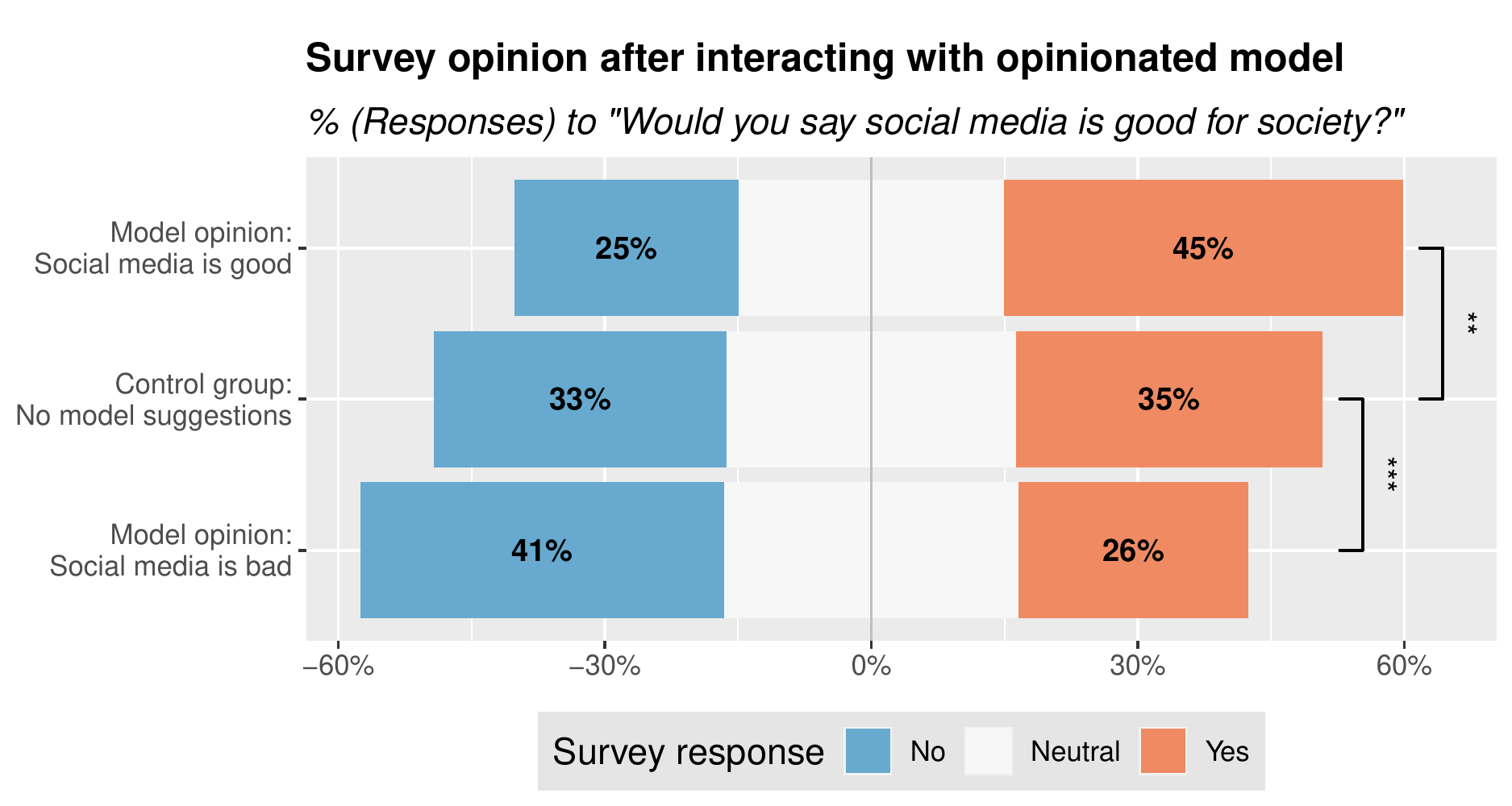}
  \end{center}
\caption{\textbf{Participants interacting with a model supportive of social media were more likely to say that social media is good for society in a later survey (and vice versa).} N\textsubscript{r}=\x{1,506} survey responses by N\textsubscript{r}=\x{1,506} participants. The y-axis indicates whether participants received suggestions from a model supportive or critical of social media during the writing task. The x-axis shows how often they said that social media was good for society (orange) or not (blue) in a subsequent attitude survey. Undecided participants are shown in white. Brackets indicate  significant opinion differences  at the **p<0.005 and ***p<0.001 level.}
\label{fig:survey-opinion}
\end{figure*}

The opinion differences in participants' writing may be due to \x{shifts} in participants' actual opinion caused by interacting with the opinionated model. We evaluate whether interactions with the language model \x{affected} participants' attitudes expressed in a post-task survey asking participants whether they thought social media was good for society. An overview of participants' answers is shown in Figure~\ref{fig:survey-opinion}.

The figure shows the frequency of different survey answers (x-axis) for the participants in each condition (y-axis).
Participants who did not interact with the opinionated models (middle row in Figure~\ref{fig:survey-opinion}) were balanced in their evaluations of social media: 33\% answered that social media is not good for society (middle, blue); 35\% said social media is good for society.
In comparison, 45\% of participants who interacted with a language model supportive of social media (top row) answered that social media is good for society. Converting participants' answers to an interval scale, this \x{difference} in opinion corresponds to an effect size of d=0.22 (p<0.001). Similarly, participants that had interacted with the language model critical of social media (bottom row) were more likely to say that social media was bad for society afterward (d=0.19, p<0.005).

\begin{figure*}
  \begin{center}
    \includegraphics[width=0.82\textwidth, trim=0cm 0cm 0cm 0cm]{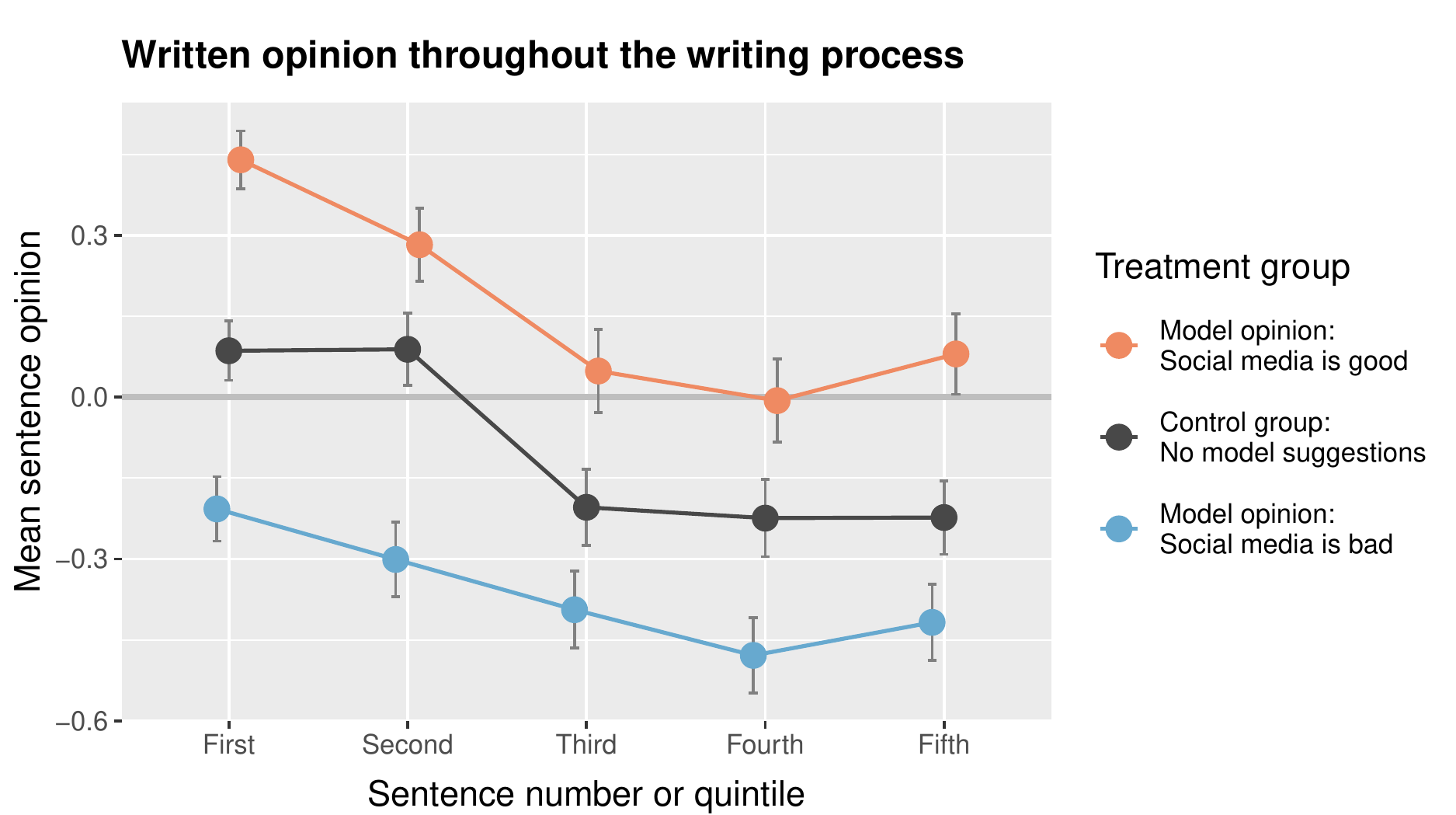}
  \end{center}
\caption{\textbf{Participants' writing was affected by the model equally throughout the writing process.} N\textsubscript{s}=9,223 sentences by N\textsubscript{p}=1,506 participants. The y-axis shows the mean opinion expressed in participants' sentences. The x-axis indicates whether the sentence was positioned earlier or later in participants' social media posts. Since most participants wrote five sentences as requested, the quintiles roughly correspond to sentence numbers.}
\label{fig:suggestions-position}
\end{figure*}

\subsubsection{Did the opinionated model gradually convince the participant?}
While we cannot ascertain the  mechanism of persuasion, our results provide further insight into how this process might have occurred. 
Figure~\ref{fig:suggestions-position} shows how participants' written opinions \x{evolved} throughout their writing process. In the control group (shown in black), participants tended to start their posts with two positive statements, followed by two more critical statements and an overall critical conclusion. Participants interacting with a model that evaluated social media positively (orange) consistently evaluated social media more favorably throughout their entire statement. Participants interacting with a model critical of social media (blue) also wrote sentences that were more critical of social media, starting with their first sentence. Similar to the control group, they were more positive at the beginning and more critical towards the end of their post, showing that the writing assistant augmented rather than replaced their narrative. 

\subsection{Were participants aware of the model's opinion and influence?}
After the writing task, we asked treatment group participants about their experience with the writing assistant. We use their answers to estimate to what extent they were aware of the model's opinion and influence.

\begin{figure*}
  \begin{center}
    \includegraphics[width=0.82\textwidth, trim=0cm 0cm 0cm 0cm]{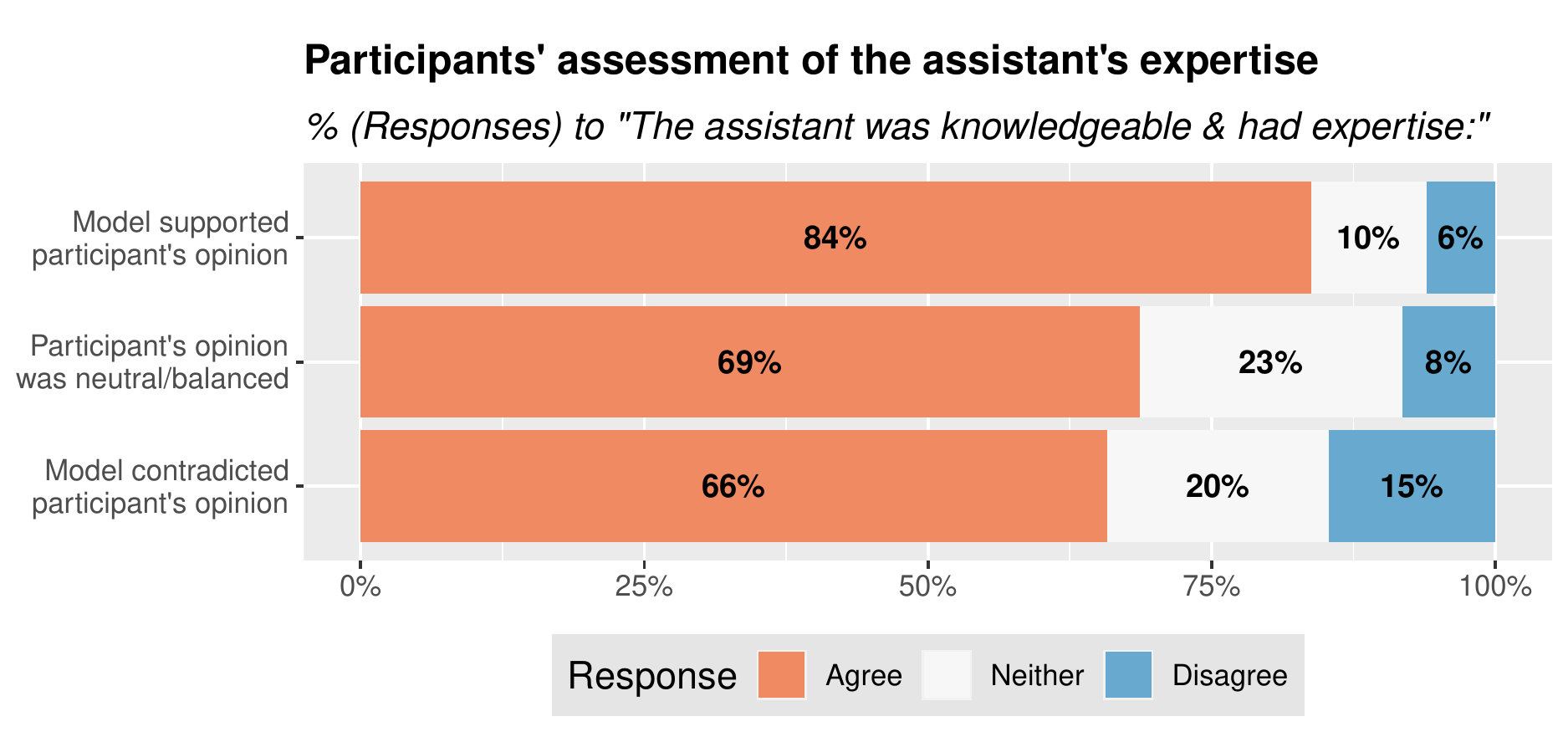}
  \end{center}
\caption{\textbf{Participants viewed the model as knowledgeable -- even if it did not share their opinion. } N\textsubscript{p}=1,000 treatment group participants. The x-axis indicates whether participants believed the language model had expertise. The y-axis indicates whether the model's opinion was aligned with participants' views. }
\label{fig:ux-expertise}
\end{figure*}

The vast majority of participants thought the language model had expertise and was knowledgeable -- even if it contradicted their personal views. As shown in Figure \ref{fig:ux-expertise}, 84\% of participants said that the assistant was knowledgeable and had expertise when the language model supported their opinion. When the model contradicted their opinion, only 15\% of participants said that it was not knowledgeable or lacked expertise.

\begin{figure*}
  \begin{center}
    \includegraphics[width=0.82\textwidth, trim=0cm 0cm 0cm 0cm]{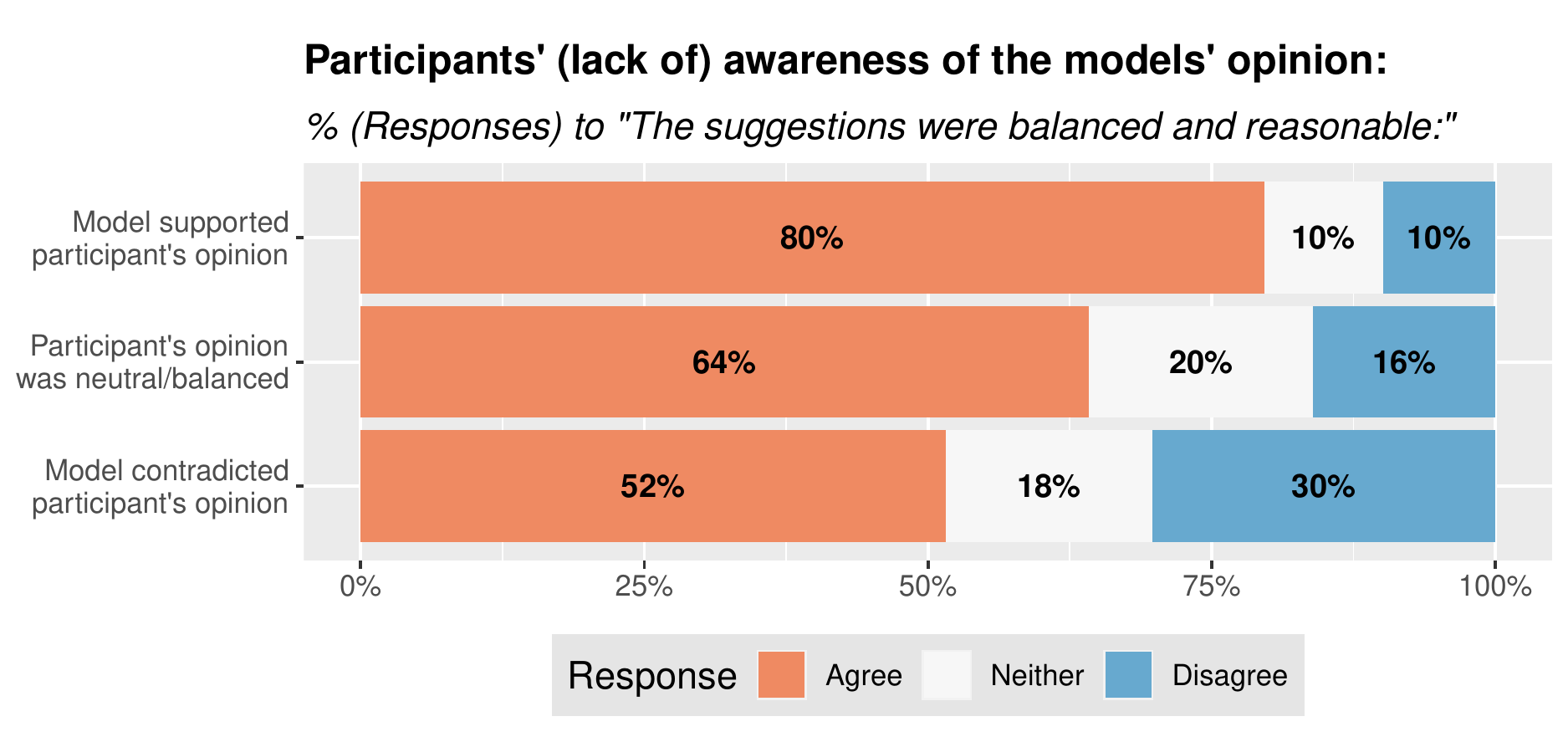}
  \end{center}
\caption{\textbf{Participants were often unaware of the model's opinion.} N\textsubscript{p}=1,000 treatment group participants. The x-axis indicates whether participants found the model's suggestions balanced and reasonable. The y-axis indicates whether the model's opinion was aligned with participants' personal views.
}
\label{fig:ux-balanced}
\end{figure*}

While the language model was configured to support one specific side of the debate, the majority of participants said that the model's suggestions were balanced and reasonable. Figure~\ref{fig:ux-balanced} shows that, in the group of participants whose opinion was supported by the model, only 10\% noticed that its suggestions were imbalanced (top row in blue). When the model contradicted participants' opinions, they were more likely (30\%) to notice its skew, but still, more than half agreed that the model's suggestions were balanced and reasonable (bottom row in orange).

\begin{figure*}
  \begin{center}
    \includegraphics[width=0.82\textwidth, trim=0cm 0cm 0cm 0cm]{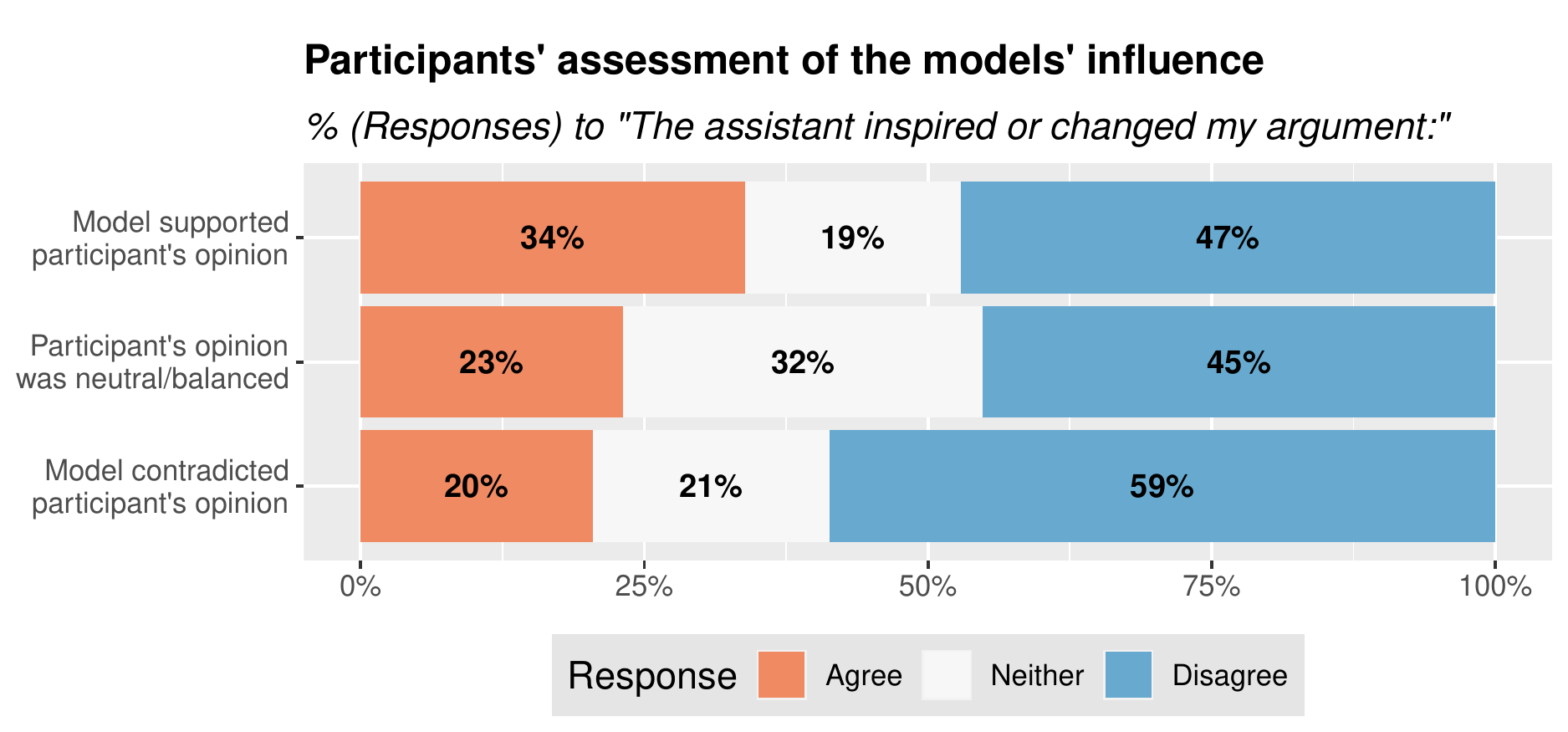}
  \end{center}
\caption{\textbf{Participants interacting with a model that supported their opinion were more likely to indicate that the model \x{affected} their argument.} N\textsubscript{p}=1,000 treatment group participants. The x-axis indicates whether participants thought that the model \x{affected} their argument. The y-axis indicates whether the model's opinion was aligned with participants' personal views.
}
\label{fig:ux-influence}
\end{figure*}

Figure~\ref{fig:ux-influence} shows that the majority of participants were not aware of the model's \x{effect} on their writing. Participants using a model aligned with their view -- and accepting suggestions more frequently -- were slightly more aware of the model's effect (34\%, top row in orange). In comparison, only about 20\% of the participants who did not share the model's opinion believed that the model influenced them. Overall, we conclude that participants were often unaware of the model's opinion and influence.




\subsection{Robustness and validation}  \label{sec:validation}
We finally validate that the experimental manipulation worked as intended and address potential concerns about experimenter demand effects.

\subsubsection{Did manipulating the models' opinion work as intended?} To validate that the prompting technique led to model output opinionated as intended, we sampled a subset of all suggestions shown to participants and asked raters in the sentence labeling task to indicate the opinion expressed in each. We found that of the full sentences suggested by the model, 86\% were labeled as supporting the intended view, and 8\% were labeled as balanced. For partially suggested sentences, that is, suggestions where the participants had already begun a sentence and the model completed it, the ratio of suggestions that were opinionated as intended dropped to 62\% (another 19\% argued that social media is both good and bad). Overall, these numbers indicate that the prompting technique guided the model to generate the target opinion with a high likelihood.

\subsubsection{Could participants have accepted the model suggestion and \x{shifted} their opinion to satisfy the experimenters?} As in all subject-based research, there is a chance that participants \x{adapted} their behavior to fit their interpretation of the study's purpose. However, we have reason to believe that demand effects do not threaten the validity of our results. When participants were asked what they perceived as the purpose of the study, most thought we were studying what people think about social media or how they use writing assistants. Only about 14\% mentioned that we might be studying the assistants' effect on people's opinions. Further, based on our post-task survey, most participants were not aware of the model's opinion and believed that the model did not \x{affect} their argument. These results suggest that participants did not adapt their views because they felt the research team expected them to.

\section{Discussion}
\x{The findings show that opinionated AI language technologies can affect what users write and think.  In our study, participants assisted by an opinionated language model were more likely to support the model's opinion in a simulated social media post} than control group participants who did not interact with a language model. Even participants who took five minutes to write their post -- ample time to write the five required sentences -- were significantly affected by the model's preferred view, showing that conveniently accepted suggestions do not explain the model's influence. Most importantly, the interactions with the opinionated model also led to opinion differences in a later attitude survey. The opinion shifts  in the survey suggest that the differences \x{in written opinion were  associated with a shift in personal attitudes. We attribute the  shifts in written opinion and post-task attitude to a new form of technology-mediated influence that we call \emph{latent persuasion} by language models.} 

\subsection{Theoretical interpretation}
\x{The literature on social influence and persuasion \cite{rashotte2007social} provides ample evidence that our thoughts, feelings, and attitudes shift due to interaction with others. Our results demonstrate that co-writing with an opinionated language model similarly shifted people's writing and attitudes. We discuss below how \emph{latent persuasion} by 
AI language technologies extends and differs from traditional social influence and conventional forms of technology-mediated persuasion \cite{simons2011persuasion}. 
We consider how the model's influence can be explained by discussing two possible vectors of influence inspired by social influence theory \cite{rashotte2007social}--informative and normative persuasion-- and a third vector of influence extending the nudge paradigm \cite{leonard2008richard, fogg2002persuasive} to the realm of opinions.}

\subsubsection{Informational influence}
\x{The language model may have influenced participants' opinions by providing new information or compelling arguments, that is, through  \emph{informational influence} \cite{myers2008social}. Some of the suggestions the language model provided may have made participants think about benefits or drawbacks of social media that they would not have considered otherwise, thus influencing their thinking. While the language model may have provided new information to writers in some cases, our secondary findings indicate that \emph{informational influence} may not fully explain the observed shifts in opinion. First,  
the model influenced participants consistently throughout the writing process. Had the language models influenced participants' views through convincing arguments, one would expect a gradual or incremental change of opinion, as has been observed for human co-writers \cite{Kimmerle2012UsingCF}. Further, our participants were largely unaware of the language model's skewed opinion and influence. The lack of awareness of the models' influence  supports the idea that the model's influence was not only through conscious processing of new information but also through the subconscious \cite{petty1986elaboration} and intuitive processes \cite{kahneman2011thinking}.}

\subsubsection{Normative influence}
The language model may have shifted participants' views through \emph{normative influence} \cite{myers2008social}. \x{Under normative influence, people adapt their opinions and behaviors based on a desire to fulfill others' expectations and gain acceptance. 
This explanation aligns with the \textit{computers are social actors} paradigm \cite{nass1994computers}, where the writing assistant may have been perceived as an independent social actor. People may have felt the need to reciprocate the language model, applying the social heuristics they apply in interactions with other humans. 
The \emph{normative influence} explanation is supported by the finding that} participants in our experiment attributed a high degree of expertise to the assistant (see Figure \ref{fig:ux-expertise}). The wider literature similarly suggests that people may regard AI systems as authoritative sources \cite{kapania2022because, logg2019algorithm, araujo2020ai}. \x{However, our experimental design presented the language model as a support tool and did not personify the assistant. An ad-hoc analysis of participants' comments on the assistant suggested that they did not feel obliged to reciprocate or comply with the models' suggestions, indicating that the strength of normative influence may have been limited. }

\subsubsection{\x{Behavioral influence}}
Large language models may \x{affect people's views by changing behaviors related to opinion formation.}
\x{The suggestions may have interrupted participants' thought processes and driven them to spend time evaluating the suggested argument \cite{Bhat2022, Buschek2021emails}.}
Similar to \emph{nudges}, the suggestions changed participants' behavior, prompting participants to consider the models' view and even accept it in their writing. 
According to self-perception theory \cite{bem1972self}, such changes in behavior may lead to changes in opinion. People who do not have strongly formed attitudes may infer their opinion from their own behavior. \x{Even participants with pre-formed opinions on the topic may have changed their attitudes by being encouraged to communicate a belief that runs counter to their own belief \cite{WAN2010162, becker2006peer}.
The finding that the model strongly influenced participants who accepted the models' suggestions frequently corroborates that some of the opinion influence has been through behavioral routes.
 The \emph{behavioral influence} route implies that the user interface and interaction design of AI language systems mediate the model's influence as they determine when, where, and how the generated opinions are presented.}

\x{We conclude that further research will be required to identify the mechanisms behind \emph{latent persuasion} by language models. Our secondary findings suggest that the influence was at least partly subconscious and not simply due to the convenience and new information that the language model provided. Rather, co-writing with the language model may have changed participants' opinion formation process on a behavioral level.}

\subsection{Implications for research and industry} 
Our results caution that interactions with opinionated language models affect users' opinions, even if unintended. 
\x{The results also show how simple it is to make models highly opinionated using accessible methods like prompt engineering.
How can researchers, AI practitioners, and policymakers respond to this finding? We believe that our results imply that we must be more careful about the opinions we build into AI language technologies like GPT-3.}

Prior work on the societal risks of large language models has warned that models learn stereotypes and biases from their training data \cite{bender2021dangers, caliskan2017semantics, garrido2021survey} that may be amplified through widespread deployments \cite{Blodgett_power_2020}. 
Our work highlights the possibility that large language models  reinforce not only stereotypes but all kinds of opinions -- from whether social media is good to whether people should be vegetarians and who should be the next president. 
Initial tools have been developed for monitoring and mitigating generated text that is discriminating~\cite{huang2019reducing, brown2020language, nozza2021honest} or otherwise offensive~\cite{askell2021general}.
We have no comparable tools for monitoring the opinions built into large language models and in the text they generate during use. A first exploration of the opinions built into GTP-3 by \citet{johnson2022ghost} suggests that the model's preferred views align with dominant US public opinion. In addition, a version of GPT trained on 4chan data led to controversy about the ideologies that training data should not contain. 
\x{We need theoretical advancements and a broader democratic discourse on what kind of opinions a well-designed model should ideally generate. }

\x{Beyond unintentional opinion shifts through carelessly calibrated models, our results raise concerns about new forms of targeted opinion influence. If large language models affect users' opinions, their influence could be used for beneficial social interventions, like reducing polarization in hostile debates or countering harmful false beliefs. However, the persuasive power of AI language technology may also be leveraged by commercial and political interest groups to amplify views of their choice, such as a favorable assessment of a policy or product.  In our experiment, we have explored the scenario of influence through a language-model-based writing assistant in an online discussion, but opinionated language models could be embedded in other applications like predictive keyboards, smart replies, and voice assistants. 
Like search engine and social media network operators \citep{knoll2016advertising}, operators of these applications may choose to monetize the persuasive power of their technology.}

\x{As researchers, we can advance an early understanding of the mechanisms and dangers of \emph{latent persuasion} through AI language technologies. 
Studies that investigate how \emph{latent persuasion} differs from other sorts of influence, how it is mediated by design factors and users' traits, and engineering work on how to  measure and guide model opinions can support product teams in reducing the risk of misuse and legislators in drafting policies that preempt harmful forms of \emph{latent persuasion}.}

\subsection{Limitations and generalizability}
\x{As appropriate for an early study, our experiment has several limitations: We only tested whether a language model affected participants' views on a single topic. We chose this topic as people had mixed views on it and were willing to deliberate. Whether our findings generalize to other topics, particularly where people hold strong entrenched opinions, needs to be explored in future studies. Further, we only looked at one specific implementation of a writing assistant powered by GPT-3. Interacting with different language models through other applications, such as a predictive keyboard that only suggests single words or an email assistant that handles entire correspondences, may lead to different influence outcomes. }

Our results provide initial evidence that language models in writing assistance tasks affect users' views. 
\x{How large is this influence compared to other types of influence, and to what extent effects persist over time, will need to be explored in future studies. 
For this first experiment, we created a \textit{strongly opinionated} model. In most cases, model opinions in deployed applications will be less definite than in our study and subject to chance variation.} However, our design also underestimates the opinion shifts that even weakly opinionated models could cause: In the experiment, participants only interacted with the model once. In contrast, people will regularly interact with  deployed models over an extended period. 
Further, in real-world settings, people will not interact with models individually, but millions will interact with the same model, and what they write with the model will be read by others. 
Finally, when language models insert their preferred views into people's writing, they increase the prevalence of their opinion in future training data, leading to even more opinionated future models.

\subsection{Ethical considerations}
\x{The harm participants incurred through interacting with the writing assistant in our study was minimal. The opinion shift was likely transient, inconsequential, and not greater than shifts ordinarily encountered in advertising on the web and TV. Yet, given the weight of our research findings, we decided to share our results with all participants in a late educational debrief: In a private message, we invited crowdworkers who had participated in the experiment and pilot studies to a follow-up task explaining our findings. We reminded participants of the experiment, explained the experimental design, and presented our results in understandable language. We also provided them with a link to a website with a nonpartisan overview of the pros and cons of social media and asked them whether they had comments about the research. 1,469 participants completed the educational debrief in a median time of 109 seconds, for which they received a bonus payment of \$0.50. We asked participants for open-ended feedback on our experiment so they could voice potential concerns. 839 participants provided open-ended comments on our experiment and results. Their feedback was exceptionally positive and is included in the Open Science Repository.}

\x{Considering the broader ethical implications of our results, we are concerned about misuse. On the one hand, we have shown how simple it is to create highly opinionated models. Our results might motivate some to develop technologies that exploit the persuasive power of AI language technology. In disclosing a new vector of influence, we face ethical tensions similar to cybersecurity researchers: On the one hand, publicizing a new vector of influence increases the chance that someone will exploit it; on the other hand, only through public awareness and discourse effective preventive measures can be taken at the policy and development level. While risky, decisions to share vulnerabilities have led to positive developments in computer safety \citep{macnish2020ethics}. We hope our results will  contribute to an informed debate and early mitigation of the risks of opinionated AI language technologies. }

\begin{acks}
We thank Yujie Shao for her support in performing the pilot studies for this project. This material is based upon work supported by the National Science Foundation under Grant No. CHS 1901151/1901329 and the German National Academic Foundation. Part of the work is funded by the Bavarian State Ministry of Science and the Arts and coordinated by the Bavarian Research Institute for Digital Transformation (bidt).
\end{acks}


\bibliographystyle{ACM-Reference-Format}
\balance
\bibliography{references}

\appendix

\end{document}